\documentclass[12pt]{article}
\usepackage{amsmath}
\usepackage{graphicx}
\usepackage{enumerate}
\usepackage{natbib}
\usepackage{url} 
\newcommand{\blind}{0}

\usepackage{xcolor, amssymb, caption}
\usepackage{bm,verbatim}

\newcommand{\bB}{{\bf B}}

\newcommand{\be}{{\bf e}}
\newcommand{\bbf}{{\bf f}}

\newcommand{\bI}{{\bf I}}

\newcommand{\bQ}{{\bf Q}}

\newcommand{\bR}{{\bf R}}
\newcommand{\hR}{{\hat{\bf R}}}

\newcommand{\bS}{{\bf S}}

\newcommand{\bx}{{\bf x}}
\newcommand{\by}{{\bf y}}

\newcommand{\rE}{{\rm E}}
\newcommand{\tr}{{\rm tr}}
\newcommand{\diag}{{\rm diag}}
\newcommand{\bSig}{\bm{\Sigma}}
\newcommand{\hlambda}{{\hat{\lambda}}}
\newcommand{\bPsi}{{\bm{\Psi}}}

\newcommand{\qed}{\hfill $\blacksquare$ \\}

\newtheorem{thm}{Theorem}
\newtheorem{lem}{Lemma}
\newtheorem{rem}{Remark}
\begin{document}

\def\spacingset#1{\renewcommand{\baselinestretch}%
{#1}\small\normalsize} \spacingset{1}


\if0\blind
{
  \title{\bf Estimating Number of Factors by Adjusted Eigenvalues Thresholding}
  \author{Jianqing Fan\thanks{Jianqing Fan is supported by NSF grants DMS-1712591 and DMS-1947097 and NIH grant R01-GM072611.
  Jianhua Guo and Shurong Zheng gratefully acknowledge National Natural Science Foundation of China (NNSFC) grant 11690012.}\hspace{.2cm}\\
    Department of Operations Research and Financial Engineering, \\
    Bendheim Center for Finance, Princeton University\\
    and \\
    Jianhua Guo$^{*}$\quad Shurong Zheng$^{*}$\\
    School of Mathematics and Statistics and KLAS, Northeast Normal University}
  \maketitle
} \fi

\if1\blind
{
  \bigskip
  \bigskip
  \bigskip
  \begin{center}
    {\LARGE\bf Estimating Number of Factors by Adjusted Eigenvalues Thresholding}
\end{center}
  \medskip
} \fi

\bigskip
\begin{abstract}
Determining the number of common factors is an important and practical topic in high dimensional factor models. The existing literatures are mainly based on the eigenvalues of the covariance matrix. Due to the incomparability of the eigenvalues of the covariance matrix caused by heterogeneous scales of observed variables, it is very difficult to give an accurate  relationship between these eigenvalues and the number of common factors.
To overcome this limitation, we appeal to the correlation matrix and show surprisingly that the number of eigenvalues greater than $1$ of population correlation matrix is the same as the number of common factors under some mild conditions.  To utilize such a relationship, we study the random matrix theory based on the sample correlation matrix in order to correct the biases in estimating the top eigenvalues and to take into account of estimation errors in eigenvalue estimation.  This leads us to propose adjusted correlation thresholding (ACT) for determining the number of common factors in high dimensional factor models, taking into account the sampling variabilities and biases of top sample eigenvalues. We also establish the optimality of the proposed methods in terms of minimal signal strength and optimal threshold.  Simulation studies lend further support to our proposed method and show that our estimator outperforms other competing methods in most of our testing cases.
\end{abstract}

\noindent%
\vfill

\newpage
\spacingset{1.5} 
\section{Introduction}
High-dimensional factor models find many applications in finance, economics, and genomics, or more generally high-dimensional data where the dependence of measurements can be attributed to a relatively small number of common factors \citep{FWZZ18}.  Determining the number of factors is an important issue in applications of factor models. The methods are typically based on eigenvalues or rank of the loading matrix. For example, \cite{Lewbel1991} and \cite{KLZ2019} obtained the number of factors by testing the rank of the loading matrix.

There are rich literatures on eigenvalues based methods for selecting the number of common factors, which have been studied from three different perspectives.  The first one is through model selection.  \cite{BaiNg2002} proposed three PC and three IC criteria by using the penalties to determine the number of common factors.
\cite{HL2007} developed an information criterion to determine the number of common factors in the general dynamic model.
\cite{LiLiShi2017} proposed the information criteria similar to \cite{BaiNg2002} to determine the number of common factors when the number of factors increases with the sample size.
\cite{SuWang2017} used the BIC information criterion to determine the number of common factors for time-varying factor models.

The second perspective is through hypothesis testing or confidence intervals.
\cite{Onatski2009} proposed a test statistic
$T_{ON}=\max\limits_{r_{\min}<i\leq r_{\max}} (\hat{\lambda}_i-\hat{\lambda}_{i+1})/(\hat{\lambda}_{i+1}-
\hat{\lambda}_{i+2})$
to test the number of common factors where $\hat{\lambda}_i$ is the $i$th largest eigenvalue of the estimated covariance matrix, and $r_{\min}$ and $ r_{\max}$ are pre-specified lower and upper
bounds of the number of common factors and estimated the number of factors by
$$
\hat{K}_{ON}=\arg\max\limits_{r_{\min}<i\leq r_{\max}} (\hat{\lambda}_i-\hat{\lambda}_{i+1})/(\hat{\lambda}_{i+1}-
\hat{\lambda}_{i+2}).
$$
\cite{Kapetanios2010} used the statistic $\tau_i(\hat{\lambda}_{i}-\hat{\lambda}_{r_{\max}+1})$ to test the number of common factors
where $\tau_i$ is the normalized constant. \cite{PanYao2008} used the Ljung-Box-Pierce portmanteau test statistic to determine the number of common factors.
Based on a confidence interval of the largest non-spiked eigenvalue of the estimated covariance matrix, \cite{CHP2017} proposed an algorithm to determine the number of common factors under the convergence regime that the dimension and the sample size tend to infinity proportionally.

The third perspective is through estimation.
\cite{Onatski2010} used the maximum eigengap to determine the number of common factors and proposed the eigenvalue difference criterion as follows:
$$
\hat{K}_{ED}=\max\{i\leq r_{\max}: \hat{\lambda}_i-\hat{\lambda}_{i+1}\geq s\},
$$
with $s$ being a given threshold. \cite{Onatski2010} stated that the difference between ED and Bai-Ng criteria is that the threshold of ED is sharp and the threshold of Bai-Ng criteria has more freedom. \cite{Wang2012} and \cite{LamYao2012} proposed to use the ratios of two adjacent eigenvalues to determine the number of factors, which estimates $K$ by
\begin{equation}\label{hatr}
\hat{K}_{ER}=\arg\max\limits_{1\leq i\leq r_{\max}}\hat{\lambda}_{i}/\hat{\lambda}_{i+1}.
\end{equation}
\cite{AH2013} proposed also  ``ER" method independently, in addition to the ``GR" method:
$$ \hat{K}_{GR}=\arg\max\limits_{1\leq i\leq r_{\max}}\log(V_{i-1}/V_i)/\log(V_i/V_{i+1}),
$$
with $V_i=\sum_{j=i+1}^{p}\hat{\lambda}_{j}$.

The aforementioned methods are all based on the eigenvalues of the covariance matrix, which is assumed to admit the sum of a low-rank matrix and a sparse matrix. Let $\bB=(b_{ij})$ be a $p\times K$ dimensional matrix with $K<p$ that represents the factor loading matrix and $\diag(\nu_{1}^2,\cdots ,\nu_{p}^2)$ be the diagonal matrix that represents the variances of idiosyncratic  noises.
Then, the covariance matrix of observed high-dimensional data is given by
\begin{equation}\label{Sigma}
\bSig=\bB\bB^T+\diag(\nu_{1}^2,\cdots ,\nu_{p}^2),
\end{equation}
where $T$ denotes the transpose of a vector or a matrix. A drawback of the covariance based methods is that it does not take into account the scales of the observed variables.
For this reason, the existing methods can easily be inconsistent.
For example, even for the simplest factor model (\ref{Sigma}) with the population covariance matrix $\bm{\Sigma}=\bB\bB^T+\diag(\underbrace{1,\cdots ,1}_{K},\nu_{K+1}^2,1\cdots ,1)$ where $\bB^T=(\bB_1^T, {\bf 0}_{K\times(p-K)})$, $\bB_1$ is of $K\times K$ dimension and ${\rm rank}(\bB_1)=K$, under some mild conditions of $\bB_1$ and $\nu_{K+1}^2$, we can show that
\begin{eqnarray}
P(\hat{K}_{ON}\geq K+1)\rightarrow1,&& P(\hat{K}_{ED}\geq K+1)\rightarrow1,\nonumber\\
P(\hat{K}_{ER}\geq K+1)\rightarrow1,&& P(\hat{K}_{GR}\geq K+1)\rightarrow1,\label{zero}
\end{eqnarray}
but in fact, the true number of common factors is $K$. The proof of (\ref{zero}) will be given in Appendix B.

The correlation matrix clearly overcomes the  scaling drawback of the covariance matrix.  The $p\times p$ dimensional correlation matrix of $\bSig$ is given by
\begin{equation}\label{cormatrix}
\bR=[\diag(\bSig)]^{-1/2}\bSig[\diag(\bSig)]^{-1/2},
\end{equation}
where $\diag(\bSig)$ is the diagonal matrix by replacing the off-diagonal elements of $\bSig$ by zeros. Using the sample correlation matrix for factor analysis will overcome the aforementioned disadvantages of using
sample covariance matrix.
In fact, when the dimension is fixed and the sample size tends to infinity,
\cite{Guttman1954}, \cite{Kaiser1960, Kaiser1961} and \cite{JW2007} (page 491) have established a lower bound:
the number of the eigenvalues satisfying $\max\{j: \hat{\lambda}_j>1, j \in \{1,\cdots,p\}\}$ is smaller than or equal to the number of the common factors.
 But the existing literatures haven't shown that they are indeed the same under certain conditions. Moreover, their estimation techniques
$$
\hat{K}_u=\max\{j: \hat{\lambda}_j>1, j \in [p]\} \quad \mbox{where} \quad [p] = \{1, \cdots, p\}
$$
can not be consistent in the high dimensional setting since
sample correlation matrices are inconsistent.
Can such a simple, tuning parameter-free method be modified so that it is consistent for high dimensional factor models?  This paper gives an affirmative answer via some high-dimensional adjustments of threshold parameters, leveraging on the random matrix theory.

The main contributions of this paper are as follows:
\begin{itemize}
\item
Firstly, we establish the concise relationship between the eigenvalues of population correlation matrices and the number of common factors, that is, give the condition under which
 \begin{equation}\label{true}
 K=\max\{j: \lambda_j(\bR)>1, j \in [p]\}
 \end{equation}
where $\lambda_1(\bR)\geq\lambda_2(\bR)\geq\cdots \geq\lambda_p(\bR)$ are the eigenvalues of correlation matrix $\bR$ and $K$ is the true number of common factors.

In factor analysis, the eigenvalues of correlation matrix are frequently used to evaluate the contributions of selected factors.
Since
$\sum_{j=1}^{p}\lambda_j(\bR)=p$, some of eigenvalues of $\bR$ are greater than 1
and the remaining eigenvalues of $\bR$ are equal to or less than 1.
It has been shown \citep{Guttman1954, Kaiser1960, Kaiser1961}  that the number of common factors is less than or equal to the number of $\bR$'s eigenvalues greater than 1. One of contributions is to show that they are indeed the same. The results presented in Table \ref{Khat} illustrate this point where $\{b_{\ell j}, \ell\in[p], j\in [K-1]\}$ are i.i.d. from the uniform distribution ${\rm U}(-1, 1)$, $\nu_{1}^2=\ldots=\nu_{p}^2=\sigma^2$, $b_{1K},\cdots ,b_{pK}$ are i.i.d. from $U(-1, 1)$ in Scenario 1 and $b_{1K}=\cdots =b_{pK}=0$ in Scenario 2. 

\begin{table}[!h]
\caption{Number of eigenvalues of $\bR$ satisfying $\lambda_j(\bR)>1$}\label{Khat}
\renewcommand{\arraystretch}{1.0}\doublerulesep 0.15pt \tabcolsep 0.2in
\begin{center}
\begin{tabular}{cc|ccc|ccc}
\hline
 &    &\multicolumn{3}{c|}{Scenario 1} &\multicolumn{3}{c}{Scenario 2}\cr
 &    &$\sigma^2=1$&2&3&$\sigma^2=1$&2&3\cr
\hline
$K$&$p$&\multicolumn{3}{c|}{{\rm rank}(\bB)=K}&\multicolumn{3}{c}{{\rm rank}(\bB)=K-1}\cr
5  &50  &5  &5  &5  &4  &4  &4\cr
   &100 &5  &5  &5  &4  &4  &4\cr
   &&&&&&\cr
10 &50  &10 &10 &10 &9  &9  &9\cr
   &100 &10 &10 &10 &9  &9  &9\cr
\hline
\end{tabular}
\end{center}
\end{table}

\item
Secondly, we propose a bias corrected estimator $\hat{\lambda}^{C}_i$ for $\lambda_i(R)$, which in general differs from the $i^{th}$ largest eigenvalue $\hlambda_i$ of sample correlation matrix and
develop a new estimator for the number of common factors as follows:
\begin{equation}\label{Khat0}
\hat{K}^C =\max\{j: \hat{\lambda}^{C}_j>s,~~j\in[r_{\max}]\}, \qquad s=1+\sqrt{p/(n-1)}
\end{equation}
under the regime $\rho_{n-1}=p/(n-1)\rightarrow \rho\in (0, \infty)$,  where $p$ is the dimension and $n$ is the sample size. Our newly proposed method $\hat{K}^C$ does not depend on any tuning parameter and  is even simpler than the eigenvalue ratio method $\hat{K}_{ER}$ or $\hat{K}_{ED}$ which involves the tuning parameter $s$.   On the other hand, by (\ref{true}), a naive method is
$$
    \hat{K}_u=\max\{j: \hat{\lambda}_j>1, j \in [p]\}.
$$
but this method overestimates $K$ when $n$ and $p$ are of the same order.


Let $p/(n-1)\rightarrow \rho$ and
\begin{eqnarray*}
\mathcal{F}(v_0)&=&\{\bR:~~{\bf R}~\mbox{is the correlation matrix (\ref{R}) of the observed vector }\\
           & &\qquad\quad\mbox{in the factor model}~(\ref{factorm})~\mbox{and}~\lambda_K(\bR)>v_0\},
\end{eqnarray*}
where $v_0$ is a positive constant, representing signal strength. We will show that the optimal lower bounds for the signal strength $v_0$ and threshold $s$ are
$$
    v_0=1+\sqrt{\rho},\quad s=1+\sqrt{p/(n-1)}
$$
in the following sense.

{\bf Minimum signal strength $v_0$}:  We will show

(i). When $v_0<1+\sqrt{\rho}$, there exists $\bR\in\mathcal{F}(v_0)$ where no method based on the eigenvalues of the sample correlation matrix can give a consistent estimate of $K$.

(ii). When $v_0=1+\sqrt{\rho}$, our method $\hat{K}^C$ can consistently estimate $K$.


 {\bf Optimal threshold $s$}.  Let $\hat{K}^C(s)$ emphasize the dependence of $\hat{K}^C$ on a general $s$.  We will prove
$$
\left\{
\begin{array}{cc}
P(\hat{K}^C(s)=K)\rightarrow 1,&\forall~\bR\in\mathcal{F}(\nu_0)~\mbox{if}~s=1+\sqrt{p/(n-1)},\\
P(\hat{K}^C(s)>K)\rightarrow 1,&\exists~\bR\in\mathcal{F}(\nu_0)~\mbox{if}~s<1+\sqrt{p/(n-1)},\\
P(\hat{K}^C(s)<K)\rightarrow 1,&\exists~\bR\in\mathcal{F}(\nu_0)~\mbox{if}~s>1+\sqrt{p/(n-1)},\\
\end{array}
\right.
$$
In other words, the threshold $s=1+\sqrt{p/(n-1)}$ is optimal. We have conducted extensive simulations to compare our method with those in \citet{BaiNg2002}, \citet{Onatski2005, Onatski2010}, \citet{LamYao2012},
\citet{AH2013}. {Simulation results show that in most of our testing cases, our estimation method outperforms the competing ones.
Even in the remaining cases considered in this paper, our estimation method has comparable performance to other competing methods.}

\item
Thirdly, we derive the asymptotic properties of the largest $K$ sample eigenvalues of the sample correlation matrix in high dimensional factor models.  This is an important contribution to random matrix theories. The results may be used for other inference problems in high dimensional factor models.
\end{itemize}

The arrangement of this paper is as follows: Section \ref{Model} reviews the factor model, defines the common factors in detail and establishes the relationship
between the number of common factors and the eigenvalues of the population correlation matrix. Section \ref{Correlation} proposes an estimation technique of the number of common factors based on a study on the random matrix theory of sample correlation matrix and demonstrates the convergence of the proposed estimator for the number of common factors.
Section \ref{Signals} investigates the optimality of the proposed estimator in high dimensional factor model.
Section \ref{Simu} presents extensive simulation results. Section \ref{ES} conducts two empirical studies.
Section \ref{Conclusion} concludes. Most of technical proofs are given in the appendix.

\section{High Dimensional Factor Model}\label{Model}
We now briefly review the factor model.  In the factor model, the observable variable $\by$ can be decomposed as
 \begin{equation}\label{factorm}
\by=\bm{\alpha}+\bB\bbf+\bm{\epsilon},
\end{equation}
where $\by=(y_1,\cdots ,y_p)^T$ is the $p$-dimensional observable vector, $\bbf=(f_1,\cdots ,f_K)^T$ is the $K$-dimensional latent factor vector, $\bm{\epsilon}=(\epsilon_1,\cdots ,\epsilon_p)^T$ is the $p$-dimensional error vector, $\bm{\alpha}$ is the $p$-dimensional intercept vector and $\bB$ is the $p\times K$ dimensional loading matrix.
Following \citet{BaiNg2002}, define the number of factors as ${\rm rank}(\bB)$.  We impose the following conditions.
\begin{itemize}
\item {\bf Condition C1:} The factors $f_1,\ldots,f_K$ are mutually independent; the factor vector $(f_1,\cdots ,f_K)$ is independent of the error vector $(\epsilon_1,\ldots,\epsilon_p)$;

\item {\bf Condition C2:} $\rE(\bbf)={\bf 0}_K, {\rm Cov}(\bbf)=\bI_K$;
\item {\bf Condition C3:} $\rE(\bm{\epsilon})={\bf 0}_p, {\rm Cov}(\bm{\epsilon})=\bPsi>{\bf 0}_{p\times p}$ where $\bPsi$ may be not diagonal (but sparse);
\item {\bf Condition C4:} $p>K$ and the loading matrix $\bB$ is of full column rank, i.e., ${\rm rank}(\bB)=K$.
\end{itemize}
Write $\bB=({\bf b}_1,\cdots ,{\bf b}_K)$ and ${\bf b}_j=(b_{1j},\cdots ,b_{pj})^T$ is a $p$-dimensional column vector for $j \in [K]$.  If there is at most one coefficient $b_{\ell j}\not=0$ with $\ell\in[p]$ for some $j\in[K]$, that is,
$f_j$ is only related to $y_j$ and not related to $y_1,..,y_{j-1},y_{j+1},\cdots ,y_p$, we can put $f_j$ in $\epsilon_j$. Thus, without loss of generality, we will define the common factor as follows:

{\bf Definition of Common Factors:}
If there are at least two coefficients $b_{\ell_1 j}, b_{\ell_2 j}\not=0$ with $\ell_1, \ell_2\in[p]$ for some $j\in[K]$, call the factor $f_j$ as a common factor.

This paper focuses on determining the number of common factors under Conditions C1-C2-C3-C4-C5 where

 {\bf Condition C5:} For every $j\in[K]$, there are at least two coefficients $b_{\ell_1 j}, b_{\ell_2 j}\not=0$ with $\ell_1, \ell_2\in[p]$.

Conditions C1-C2-C3 have been frequently imposed \citep{BaiNg2002, JW2007}.  They are related to identifiability and moment conditions. Although they are often used, Conditions C4-C5 are not explicitly written.
Condition C4 shows that $\bB$ is of full column rank, that is, ${\rm rank}(\bB)=K$. Condition C5 shows that every factor $f_j$ has an impact on at least two observed variables.

By definition (\ref{factorm}) of the factor model, we have $$\bSig={\rm Cov}(\by)=\bB\bB^T+\bPsi.$$ Let the $(j, j)$ entry of $\bSig$ be $\sigma_{jj}$ for $j \in [p]$.
Then by (\ref{cormatrix}), the population correlation matrix of $\by$ in the factor model is
\begin{equation}\label{R}
\bR=\bQ\bQ^T=\bQ_1\bQ_1^T+\bQ_2\bQ_2^T,
\end{equation}
where $\diag(\bSig)=\diag(\sigma_{11},\cdots ,\sigma_{pp})$ and
 \begin{equation}\label{Q}
 \begin{array}{lll}
 &&\bQ=[\diag(\bSig)]^{-1/2}(\bB, \bPsi^{1/2})=(\bQ_1, \bQ_2),\\
 &&\bQ_1=[\diag(\bSig)]^{-1/2}\bB,\quad\bQ_2=[\diag(\bSig)]^{-1/2}\bPsi^{1/2}.
 \end{array}
 \end{equation}
In fact, $\bQ_1\bQ_1^T$ and $\bQ_2\bQ_2^T$ include the information of the factors $f_1,\cdots ,f_K$ and errors $\epsilon_1,\cdots ,\epsilon_p$ on $\by$, respectively.
Let $\|{\bf M}\|_{F}$ denote the Frobenius norm of a matrix or a vector ${\bf M}$ and $\|{\bf M}\|=\sqrt{\lambda_1({\bf M}{\bf M}^T)}$ the operator norm.  The following theorem shows how to determine the number of factors from the population correlation matrix.

\begin{thm}\label{thm0}
Under Conditions C1-C2-C3-C4-C5, if $\|[\diag(\bSig)]^{-1}\bPsi\|\leq1$, we have
$$
    \lambda_j(\bR)\leq1,\quad j=K+1,\ldots,p.
$$
In addition, we have
\begin{equation}\label{K}
K=\max\{j: \lambda_j(\bR)>1, j \in [p]\},
\end{equation}
when $p$ is large enough and there exists three non-negative constants $\delta_1>\delta_2+\delta_3\geq0$, $\delta_3<0.5$ satisfying
\begin{eqnarray}
&&\|[\diag(\bSig)]^{-1/2}\bB\|^2_{F}=O(p^{\delta_1}),\quad K=O(p^{\delta_3}),\nonumber\\
&&\|\bB^T[\diag(\bSig)]^{-1}\bB\|\cdot\|\{\bB^T[\diag(\bSig)]^{-1}\bB\}^{-1}\|=O(p^{\delta_2}),\label{Con}\\
&&\|[\diag(\bSig)]^{-1}\bPsi\|\leq1.\nonumber
\end{eqnarray}
\end{thm}

Theorem \ref{thm0} gives a sufficient condition to ensure that the number of $\bR$'s eigenvalues greater than 1 is equal to the number of common factors.  Note that \eqref{Con} imposes a restriction on the condition number of the matrix $\bB^T[\diag(\bSig)]^{-1}\bB$. Without loss of generality, assume $\diag(\bSig)=\bI_p$. Then $\bR=\bB\bB^T+\bPsi$ and the conditions (\ref{Con}) become
$$
K=O(p^{\delta_3}),~\|\bPsi\|\leq1,~\|\bB\|^2_{F}=O(p^{\delta_1}),~\|\bB^T\bB\|\cdot\|(\bB^T\bB)^{-1}\|=O(p^{\delta_2}).
$$

\section{Properties of sample correlation matrix under factor model}\label{Correlation}

Let $\by_1, \ldots, \by_n$ be an i.i.d. sample of size $n$ from (\ref{factorm}):
$$
    \by_i=\bm{\alpha}+\bB\bbf_i+\bm{\epsilon}_i,~i\in[n] = \{1,\cdots,n\}.
$$
Then the sample covariance matrix and sample correlation matrix are
\begin{equation}\label{hatSigma}
\begin{array}{l c l}
\hat{\bm{\Sigma}}_n & =& n^{-1}\sum\limits_{i=1}^{n}(\by_i-\bar{\by})(\by_i-\bar{\by})^T,\\
\hat{\bR} & = & [\mbox{diag}(\hat{\bm{\Sigma}}_n)]^{-1/2}\hat{\bm{\Sigma}}_n[\mbox{diag}(\hat{\bm{\Sigma}}_n)]^{-1/2},
\end{array}
\end{equation}
where $\bar{\by}=n^{-1}\sum_{i=1}^{n}\by_i$ is the sample mean.
Let the empirical spectral distributions ({\it ESD}) of $\hR$ and $\bR$ be $F_n(t)$ and $H_{p-K}(t)$ as follows:
\begin{equation}\label{Fnt}
F_n(t)=\frac{1}{p-K}\sum_{j=K+1}^{p}1{(\lambda_j(\hR)\leq t)},~~H_{p-K}(t)=\frac{1}{p-K}\sum_{j=K+1}^{p}1{(\lambda_j(\bR)\leq t)},
\end{equation}
for any real number $t$ with $1(\cdot)$ being an indicator function.

\subsection{Spectral properties of sample correlation matrix}
{In order to estimate the number of common factors, we first derive some fundamental results in random matrix theories: the Stieltjes equation of the limiting spectral distribution ({\it LSD}) $F(t)$ of the ESD $F_n(t)$ and
the almost sure convergence of sample spiked eigenvalues $\lambda_1(\hR),\cdots ,\lambda_K(\hR)$ of $\hR$ . There are some existing literatures on the spectral properties of $\hR$ when $\bR$ is of special structures. Bao, Pan and Zhou (2011) derived the Tracy-Widom law of the maximum eigenvalue of $\hR$ as $\bR=\bI_p$ and $p/n\rightarrow \rho\in (0, \infty)$.
\cite{el2007spectral} established the LSD of $\hR$ for the elliptical distribution as $p/n\rightarrow \rho\in (0, \infty)$ with the bounded spectral norm $\|\bR\|$.
\cite{GHPY2017} obtained the central limit theorem of $\hR$ for the case $\bR=\bI_p$ and $p/n\rightarrow \rho\in (0, \infty)$.
However, for the general factor model (\ref{factorm}), the population correlation matrix is not $\bI_p$. Theorem \ref{thm1} below gives the Stieltjes equation of the LSD of $\hR$ for general case. For convergence of sample spiked eigenvalues $\lambda_1(\hR),\cdots ,\lambda_K(\hR)$, we have not found the related literatures and Theorem \ref{thm2} below fills the void.

In order to derive Theorems \ref{thm1}-\ref{thm2}, additional assumptions are needed.

{\bf Assumption (a)}. Letting $\bx_i=(x_{1i},\cdots ,x_{p+K,i})^T=(f_{1i},\cdots, f_{Ki}, e_{1i},\cdots, e_{pi})^T$, $(e_{1i},\cdots ,e_{pi})=(\epsilon_{1i},\cdots ,\epsilon_{pi})\bm{\Psi}^{-1/2}$,
$\{x_{ji}, j\in[p+K], i\in[n]\}$ are independent random variables satisfying:
 \begin{equation}\label{(Linde)}
 \frac1{n(p+K)\eta_n^4}\sum_{j=1}^{p+K}\sum_{i=1}^n\rE|x_{ji}^4|1{(|x_{ji}|>\eta_n\sqrt{n})}\to0,
 \end{equation}
  where $0<K\eta_n\to0$ and $K\eta_n\log n\rightarrow+\infty$.

{\bf Assumption (b)}. $\sup\limits_{j \in [p+K]}\rE(|x_{j1}|^{6+\delta_0})$ is bounded for all $p, K$ for some $\delta_0 > 0$.

{\bf Assumption (c)}. The ratio of dimension to sample size $\rho_n=p/n\to \rho\in(0, \infty)$ as $n\to\infty$.

{\bf Assumption (d)}. The number of common factors satisfies $K=o(p^{1/6})$.

{\bf Assumption (e)}. $\|[\diag(\bSig)]^{-1}\bPsi\|\leq1$ and the limiting spectral distribution $H(t)$ of the ESD $H_{p-K}(t)$ from the eigenvalues $\lambda_{K+1}(\bR),\cdots ,\lambda_p(\bR)$ of $\bR$ exists.

\begin{rem}\label{rem41}The assumption (\ref{(Linde)}) is the Lindeberg condition.
 By Theorem \ref{thm0}, it is known that $\lambda_j(\bR)\leq 1$ for $j=K+1,\cdots ,p$ if $\|[\diag(\bSig)]^{-1}\bPsi\|\leq1$. Thus, the support set of $H(t)$ is in $[0, 1]$.
\end{rem}

\begin{lem}\label{lemmax}
For the high dimensional factor model (\ref{factorm}) satisfying Conditions C1-C2-C3-C4-C5, under Assumptions (a)-(b)-(c)-(d)-(e), we have
$$
\max\limits_{j \in [p]}|\hat{\sigma}_{jj}-1|=o_{a.s.}(1),
$$
where $\hat{\sigma}_{jj}=n^{-1}\sum_{i=1}^n\be_j^T\bQ(\bx_i-\bar{\bx})(\bx_i-\bar{\bx})^T\bQ^T\be_j$ with $\bar{\bx}=n^{-1}\sum_{i=1}^n\bx_i$, $\bQ$ being defined in (\ref{Q}) and $\be_j$ is the $j$th column of $\bI_p$.
\end{lem}

The proof of Lemma~\ref{lemmax} is given in Appendix D.  As we impose weak moment conditions, we need to use the truncation tricks and hence the proof is somewhat lengthy.  For $z\in\cal{C}^{+}$, let the Stieltjes transform be
 \begin{eqnarray}
 m_n(z)&=&(p-K)^{-1}\sum_{j=K+1}^p(\lambda_j(\hR)-z)^{-1}=\int\frac{1}{t-z}dF_n(t),\nonumber\\
 \underline{m}_n(z)&=&\int\frac{1}{t-z}d\underline{F}_n(t)=-(1-\rho_{K,n-1})z^{-1}+\rho_{K,n-1} m_n(z),\label{mnz}\\
   m(z)&=&\int\frac{1}{t-z}dF(t),~~\underline{m}(z)=\int\frac{1}{t-z}d\underline{F}(t)=-(1-\rho)z^{-1}+\rho m(z),\nonumber
 \end{eqnarray}
 where $F_n(z)$ is defined in (\ref{Fnt}), $\rho_{K,n-1}=(p-K)/(n-1)$, $\underline{F}_n(x)=(1-\rho_{K,n-1})1(x>0)+\rho_{K,n-1} F_n(x)$, $\underline{F}(x)=(1-\rho)1(x>0)+\rho F(x)$ and $\cal{C}^{+}$ denotes the upper plane of the two-dimensional complex space. Then $\underline{m}(z)$ and $m(z)$ satisfy the equations \eqref{um} and \eqref{psi}.

\begin{thm}\label{thm1} For the high dimensional factor model (\ref{factorm}) satisfying Conditions C1-C2-C3-C4-C5 and Assumptions (a)-(b)-(c)-(d)-(e), we have
$$
|m_n(z)-m(z)|=o_{a.s.}(1),\quad |\underline{m}_n(z)-\underline{m}(z)|=o_{a.s.}(1),
$$
\begin{equation}\label{um}
z=-\frac{1}{\underline{m}(z)}+\rho\int\frac{tdH(t)}{1+t\underline{m}(z)}=-\underline{m}^{-1}(z)\psi(-\underline{m}^{-1}(z)),
\end{equation}
where $z\in\cal{C}^{+}$  and
\begin{equation}\label{psi}
\quad\psi(x)=1+\rho\int\frac{t}{x-t}dH(t).
\end{equation}
\end{thm}

\subsection{Bias correction of sample eigenvalues}
Let $\hlambda_j=\lambda_j(\hR)$ and $\lambda_j=\lambda_j(\bR)$ for $j\in[p]$. For any given $j$ , define
 \begin{eqnarray*}
 m_{n, j}(z)&=&(p-j)^{-1}[\sum_{\ell=j+1}^p(\hlambda_{\ell}-z)^{-1}+((3\hlambda_j+\hlambda_{j+1})/4-z)^{-1}],\\
 \underline{m}_{n, j}(z)&=&-(1-\rho_{j, n-1})z^{-1}+\rho_{j, n-1}m_{n, j}(z),\\
 \end{eqnarray*} with $\rho_{j, n-1}=(p-j)/(n-1)$. Let the corrected eigenvalue of $\hlambda_j$ be
 $$
 \hlambda_j^C=-\frac{1}{\underline{m}_{n, j}(\hlambda_j)},~j\in[r_{\max}].
 $$

The following theorem, whose proof is given in Appendix I, shows that the corrected empirical eigenvalues are consistent.

\begin{thm}\label{thm2} For the high dimensional factor model (\ref{factorm}) satisfying Conditions C1-C2-C3-C4-C5 and Assumptions (a)-(b)-(c)-(d)-(e),  for $j\in [K]$, if $\lambda_j\geq\lambda_{K+1}(\bR)(1+\sqrt{\rho})+ \delta$ for some $\delta > 0$,
\begin{equation}\label{TRUEE}
\frac{\hat{\lambda}_j^C}{\lambda_j}=1+o_p(1)\quad\mbox{and}\quad\frac{\hlambda_j}{\lambda_j}=\psi(\lambda_j)+o_p(1),
\end{equation}
In particular, if in addition $\lambda_j$ is bounded for $j\in[K]$, we have
$$
\hat{\lambda}_j^C=\lambda_j+o_p(1)\quad\mbox{and}\quad\hlambda_j=\lambda_j\psi(\lambda_j)+o_p(1).
$$
\end{thm}
\begin{rem}\label{rem1} By Remark \ref{rem41} and (\ref{K}), we have $\lambda_j>1$ for $j \in [K]$ under the conditions of Theorem \ref{thm0} and the support of $H(t)$ being in $[0, 1]$.
By (\ref{psi}) and (\ref{TRUEE}), if $\lambda_j>\lambda_{K+1}(1+\sqrt{\rho})$ and is bounded,  we have
$$
\hlambda_j-\lambda_j=\lambda_j\psi(\lambda_j)-\lambda_j=\rho\int\frac{\lambda_jt}{\lambda_j-t}dH(t)+o_p(1),~j \in [K].
$$
In other words, the sample eigenvalue $\hlambda_j$ is not a consistent estimator of $\lambda_j$ for $j \in [K]$.
This is due to the inconsistency of the high dimensional sample correlation matrix. On the other hand, from (\ref{TRUEE}), we show that the corrected eigenvalue $\hat{\lambda}^C_j$ is consistent for $j \in [K]$.
\end{rem}

\section{Minimum signals and optimal threshold}\label{Signals}


We will adopt the notation and estimator defined in the introduction.  Our aim is to find minimal signal strength $v_0$ for consistent estimation of the number of factors and to give the optimal threshold level for our estimator.

\subsection{Minimal signal strength}
The following theorem shows the minimal signal strength.

\begin{thm}$(${\rm \bf Minimal signal strength $v_0$}$)$.
 For the high dimensional factor model (\ref{factorm}) satisfying Conditions C1-C2-C3-C4-C5 and Assumptions (a)-(b)-(c)-(d)-(e),  and for any estimation method $\hat{K}_{any}$ of the number of common factors by detecting the difference between $\{\lambda_j(\hR), j\in[K]\}$ and $\{\lambda_j(\hR), j=K+1, \cdots ,p\}$, it holds that
 $$
\limsup_{n\rightarrow\infty}\inf_{\bR\in\mathcal{F}(v_0)} P(\hat{K}_{any}=K)<1,
 $$
 if $v_0<1+\sqrt{\rho}$.
\end{thm}
{\bf Proof}. Let us take $\epsilon$ such that $(1-\epsilon)(1+\sqrt{\rho})>\max\{1, v_0\}$  and  $\bR\in\mathcal{F}(v_0)$ such that $\bR = \bm{\Sigma}=\diag(\bR)$  and
$$
\lambda_p(\bR)=\cdots=\lambda_{K+1}(\bR)=1-\epsilon<\lambda_{K}(\bR)=(1-\epsilon)(1+\sqrt{\rho}).
$$
Then by (\ref{TRUEE}), we have $\lambda_{K}(\hR)=(1-\epsilon)(1+\sqrt{\rho})^2+o_{p}(1)$ because $H(t)$ is the limit of the empirical distribution function of $\{1-\epsilon,\cdots ,1-\epsilon\}$.
By (S.46) in the supplementary material, we have $|\lambda_{K+1}(\hR)-\lambda_{K+1}(\bS_n)|=o_{a.s.}(1)$.
By Theorem 1.1 of \cite{BaikSilverstein2006}, we have $(1-\epsilon)^{-1}\lambda_{K+1}(\bS_n)=(1+\sqrt{\rho})^2+o_{a.s.}(1)$.
Thus we have
\begin{equation}\label{ex1}
\lambda_{K+1}(\hR)=(1-\epsilon)(1+\sqrt{\rho})^2+o_{a.s.}(1).
\end{equation}
Hence, when $n, p$ are large enough, $\lambda_{K}(\hR)$ and $\lambda_{K+1}(\hR)$ will be indifferentiable. That is, when $n, p$ are large enough,
the difference between $\lambda_{K}(\hR)$ and $\lambda_{K+1}(\hR)$ can't be detected by any method.  

\begin{rem}The above theorem shows that
$v_0=1+\sqrt{\rho}$ is the minimal signal strength.
Thus, throughout the rest of the paper, we will consider the estimation method in the set of the correlation matrix $\bR$ as follows:
\begin{eqnarray*}
\mathcal{F}(1+\sqrt{\rho})&=&\{\bR:~~{\bf R}~\mbox{is the correlation matrix (\ref{R}) of the observed vector}\\
           & &\qquad\quad\mbox{in the factor model}~(\ref{factorm})~\mbox{and}~\lambda_K(\bR)>1+\sqrt{\rho}\}.
\end{eqnarray*}
\end{rem}

\subsection{Optimal threshold}
Recall our estimation method,
\begin{equation}\label{Kt4}
\hat{K}^C(s)=\max\{j: \hat{\lambda}^{C}_j>s, j\in [r_{\max}]\},
\end{equation}
where $r_{\max}$ is a pre-specified positive integer and the maximum of the empty set is defined as 0. The following theorem establishes the optimal bound of the threshold $s$.
\begin{thm}\label{thmL}
For the high dimensional factor model (\ref{factorm}) satisfying Conditions C1-C2-C3-C4-C5 and Assumptions (a)-(b)-(c)-(d)-(e),  we have
 \begin{eqnarray}
\limsup_{n\rightarrow\infty}\sup\limits_{\bR\in\mathcal{F}(1+\sqrt{\rho})} P(\hat{K}^C(s)>K)=1,~~\mbox{if } s<1+\sqrt{\rho},\label{U1}\\
\limsup_{n\rightarrow\infty}\sup\limits_{\bR\in\mathcal{F}(1+\sqrt{\rho})} P(\hat{K}^C(s)<K)=1,~~\mbox{if } s>1+\sqrt{\rho},\label{U2}
 \end{eqnarray}
 where $s$ doesn't depend on $n$ and $p$.
\end{thm}
{\bf Proof.} To (\ref{U1}), let $(1-\epsilon)(1+\sqrt{\rho})>s$ and $\bR\in\mathcal{F}(1+\sqrt{\rho})$ satisfy
$$
\lambda_p(\bR)=\ldots=\lambda_{K+1}(\bR)=1-\epsilon
<1+\sqrt{\rho}<\lambda_{K}(\bR).
$$
By (\ref{um}) and (\ref{TRUEE}), we have
$\lambda_{K+1}(\hR)=\hlambda_{K+1}^C+\rho\frac{\hlambda_{K+1}^C(1-\epsilon)}{\hlambda_{K+1}^C-(1-\epsilon)}+o_p(1),$
because $H(t)$ is the limit of the empirical distribution function of $\{1-\epsilon,\cdots ,1-\epsilon\}$. It then follows from $\lambda_{K+1}(\hR)=(1-\epsilon)(1+\sqrt{\rho})^2+o_p(1)$ that
$$
\hlambda_{K+1}^C=(1-\epsilon)(1+\sqrt{\rho})+o_p(1).
$$
 That is,
$\hlambda_{j}^C\geq(1-\epsilon)(1+\sqrt{\rho})+o_p(1),~j \in [K+1].$
By using $(1-\epsilon)(1+\sqrt{\rho})>s$, we conclude that
$$
\limsup_{n\rightarrow\infty}\sup_{\bR\in\mathcal{F}(1+\sqrt{\rho})}
P(\hat{K}^C(s)>K)=1,
$$
when $s<1+\sqrt{\rho}$.

To prove (\ref{U2}), let $\bSig=\bR\in\mathcal{F}(1+\sqrt{\rho})$ satisfy
$$
\lambda_p(\bR)\leq\cdots\leq\lambda_{K+1}(\bR)\leq1<1+\sqrt{\rho}<\lambda_{K}(\bR)<s\leq\cdots \leq\lambda_1(\bR).
$$
By (\ref{TRUEE}) and $\lambda_K(\bR)<s$, we have
$\hlambda_{K}^C=\lambda_K(\bR)+o_p(1)<s+o_p(1)$
which means $\hlambda_{j}^C<s,~j=p,\cdots ,K$ in probability. Thus
$$
\limsup_{n\rightarrow\infty}\sup\limits_{\bR\in\mathcal{F}(1+\sqrt{\rho})} P(\hat{K}^C(s)<K)=1,
$$
if $s>1+\sqrt{\rho}$.  \qed

Theorem \ref{thmL} shows that the choices $s<1+\sqrt{\rho}$ and $s>1+\sqrt{\rho}$ are not  optimal for the threshold parameter $s$ in our estimation method. The following
theorem will show that $s=1+\sqrt{\rho}$ is optimal.

\begin{thm}\label{thmE}
For the high dimensional factor model (\ref{factorm}) satisfying Conditions C1-C2-C3-C4-C5 and Assumptions (a)-(b)-(c)-(d)-(e), for $\bR\in\mathcal{F}(1+\sqrt{\rho})$, we have when $s=1+\sqrt{\rho}$,
 \begin{equation}\label{Kt3}
 P(\hat{K}^C(s)=K)\rightarrow1.
 \end{equation}
\end{thm}
{\bf Proof.} For any $\bR\in\mathcal{F}(1+\sqrt{\rho})$, we have
$$
\lambda_p(\bR)\leq\cdots\leq\lambda_{K+1}(\bR)\leq 1<1+\sqrt{\rho}+\epsilon_0<\lambda_{K}(\bR)\leq\cdots \leq\lambda_1(\bR),
$$
for a very small positive constant $\epsilon_0$.
By (\ref{um}), we have
$\lambda_{K+1}(\hR)=\hlambda_{K+1}^C+\rho\int\frac{\hlambda_{K+1}^Ct}{\hlambda_{K+1}^C-t}dH(t)+o_p(1).$
Thus, we have
\begin{equation}\label{Kt1}
\hlambda_{K+1}^C\leq1+\sqrt{\rho}+o_p(1),
\end{equation}
because of $\lambda_{K+1}(\hR)\leq(1+\sqrt{\rho})\psi(1+\sqrt{\rho})+o_p(1)$ by Lemma S.6 in the supplementary material. By (\ref{TRUEE}), we have
\begin{equation}\label{Kt2}
\hlambda_{j}^C\geq1+\sqrt{\rho}+\epsilon_0+o_p(1),~j \in [K].
\end{equation}
Thus, by (\ref{Kt1}) and (\ref{Kt2}),
when $s=1+\sqrt{\rho}$, we have
$$
\lim_{n\rightarrow\infty}P(\hat{K}^C(s)=K)=1.
$$
\vskip 0.5cm
{\bf Summary of Method:}  We propose
{$$
\hat{K}=\max\{j: \hat{\lambda}^{C}_j>1+\sqrt{\rho_{n-1}}, j\in [r_{\max}]\},\quad $$}
where $\rho_{n-1}=p/(n-1)$.  This is a simple and tuning free method.

\section{Simulation studies}\label{Simu}
We evaluate the finite-sample performance of the proposed method by simulation studies. Because our proposed estimating method is based on the adjusted correlation thresholding,  we will label the proposed estimating method as ``ACT". We compare our $ACT$ method with 13 existing methods:
``$PC_1$", ``$PC_2$",``$PC_3$",``$IC_1$", ``$IC_2$" and ``$IC_3$" in \citet{BaiNg2002}, ``$ON_1$", ``$ON_2$" and ``$ON_3$" in Onasti (2005), ``$NON$" in \citet{Onatski2010},
``$ER$" and ``$GR$" in \citet{AH2013}. Due to the similarity of simulation results, we only present $PC_3$, $IC_3$, $ON_2$, $ER$, $GR$ and $ACT$.
The sample sizes are taken to be $n=150, 300$ and the dimension is $p=100, 300, 500, 1000$.
Recall the factor model $\by=\bB\bbf+\bm{\epsilon}$ in (\ref{factorm}).
For the Gaussian population, assume that $\epsilon_i$'s are iid from $N(0, \nu_i^2)$ and $f_1,\cdots ,f_K$ are iid from $N(0, 1)$.
For the uniform population, assume that $\epsilon_i$ are iid from Unif$(0, 2\sqrt{3\nu_i^2})$ and $f_1,\cdots ,f_K$ are iid from Unif$(0, 2\sqrt{3})$.
We set the true number of common factors $K=5$.
For every case, we conduct 1000 replications to summarize the empirical percentages of {true estimation,} overestimation, underestimation of the number of common factors,
and the average number of common factors. We consider the following four cases for the factor loading matrix $\bB=(b_{\ell j})_{\ell\in[p], j\in[K]}$.  They can be verified to satisfy the imposed conditions, in particular, those in Theorem~\ref{thm0}.
\begin{itemize}
\item
[\bf Case] 1: Let $b_{\ell j}=\sqrt{3p^{-1/2}}$ for $\ell, j\in[K]$
and $b_{\ell j}=a_{\ell j}\sqrt{3(p-j)^{-1}}$ for $\ell\in\{K+1,\cdots ,p\}, j\in[K]$ and $a_{\ell j}=-1$ if $\ell=K j$ or $a_{\ell j}=1$ if $\ell \not=K j$.
Assume that $\nu_{1}^2=\cdots =\nu_{p}^2=0.55^2$. The model is from \cite{Harding2013}.
\item
[\bf Case] 2: Let $b_{\ell j}$ be iid from $N(0, 1)$ and $\nu_{1}^2,\cdots ,\nu_{p}^2$ be iid from Unif$(0, 180)$.
\item
[\bf Case] 3: Let $b_{\ell j}$ be iid from $N(0, 1)$ and $\nu_{1}^2=\cdots =\nu_{p}^2=36$. The model is used in \cite{BaiNg2002} and \cite{Onatski2010}.
\item
[\bf Case] 4: Let $b_{jj}=1$, $b_{\ell j}$ be iid from $N(0, 0.04)$ for $j\not=\ell$ and $\nu_{1}^2,\cdots ,\nu_{p}^2$ be iid from Unif$(0, 5.5)$.
\end{itemize}
The simulation results for Cases 1--4 with $n=300$ are presented respectively in Tables \ref{table2}--\ref{table5}.
The results for the cases with $n = 150$ are similar and are omitted.
From these tables, we can see that except for very few settings, ``ACT" behaves very well for almost all parameter setups. Even for these few settings, the percentiles of true estimation of ``ACT"
are also similar to those of ``ER" and ``GR".

\begin{table}[htbp]
\caption{Percentages of the estimated number of common factors for Case 1 with $n=300$ in 1000 simulations:
``TRUE", ``OVER" and ``UNDER" truly estimates, overestimates and underestimates the number of common factors, respectively. ``AVE" is the average of the estimated number of common factors.}
\renewcommand{\arraystretch}{0.56}\doublerulesep 0.01pt\tabcolsep 0.18in
{\small\begin{center}
\begin{tabular}{cc|cccccc}
\hline
\hline
$p$&   &$PC_3$ &$IC_3$ &$ON_2$ &$ER$ &$GR$ &ACT \cr
\hline
   &  &\multicolumn{6}{c}{Gaussian population}\cr
\hline
100 &TRUE    &99.8       &87.6      &100      &41.8    &75.7   &100\cr
    &OVER    &0          &0         &0        &0       &0      &0\cr
    &UNDER   &0.2        &12.4      &0        &58.2    &24.3   &0\cr
    &AVE     &5          &4.88      &5        &2.87    &4.37   &5\cr
\hline
300 &TRUE    &92.0       &55.9      &99.9     &4.2     &8.7    &100\cr
    &OVER    &0          &0         &0.1      &0       &0      &0\cr
    &UNDER   &8,0        &44.1      &0        &95.8    &91.3   &0\cr
    &AVE     &4.92       &4.56      &5        &2.09    &2.56   &5\cr
\hline
500 &TRUE    &0          &0         &100      &0       &0.2    &99.6\cr
    &OVER    &0          &0         &0        &0       &0      &0.4\cr
    &UNDER   &100        &100       &0        &100     &99.8   &0\cr
    &AVE     &3.88       &3.52      &5        &1.78    &1.96   &5\cr
\hline
1000&TRUE    &0          &0         &79.1     &0       &0     &89.0\cr
    &OVER    &0          &0         &0        &0       &0     &1.9\cr
    &UNDER   &100        &100       &20.9     &100     &100   &9.1\cr
    &AVE     &1.81       &1.33      &4.79     &1.34    &1.38  &4.93\cr
\hline
    &  &\multicolumn{6}{c}{Uniform population}\cr
\hline
100 &TRUE    &99.9      &90.3       &99.9     &44.7    &81.2   &100\cr
    &OVER    &0         &0          &0.1      &0       &0      &0\cr
    &UNDER   &0.1       &9.7        &0        &55.3    &18.8   &0\cr
    &AVE     &5         &4.9        &5        &2.97    &4.52   &5\cr
\hline
300 &TRUE    &93.3      &62.0       &100      &4.2     &9.0      &100\cr
    &OVER    &0         &0          &0        &0       &0      &0\cr
    &UNDER   &6.7       &38.0       &0        &95.8    &91.0     &0\cr
    &AVE     &4.93      &4.62       &5        &2.07    &2.55   &5\cr
\hline
500 &TRUE    &0         &0          &99.9     &0.1     &0.3    &99.2\cr
    &OVER    &0         &0          &0.1      &0       &0      &0.8\cr
    &UNDER   &100       &100        &0        &99.9    &99.7   &0\cr
    &AVE     &3.92      &3.55       &5        &1.75    &1.97   &5.01\cr
\hline
1000&TRUE    &0         &0          &83.3     &0       &0      &89.8\cr
    &OVER    &0         &0          &0.1      &0       &0      &1.5\cr
    &UNDER   &100       &100        &16.6     &100     &100    &8.7\cr
    &AVE     &1.82      &1.29       &4.84     &1.32    &1.37   &4.93\cr
\hline
\end{tabular}
\end{center}\label{table2}}
\end{table}

\begin{table}[htbp]
\caption{Percentages of the estimated number of common factors for Case 2 with $n=300$ in 1000 simulations:
``TRUE", ``OVER" and ``UNDER" truly estimates, overestimates and underestimates the number of common factors, respectively. ``AVE" is the average of the estimated number of common factors.}
\renewcommand{\arraystretch}{0.56}\doublerulesep 0.01pt\tabcolsep 0.18in
\begin{center}
\begin{tabular}{cc|cccccc}
\hline
\hline
$p$&   &$PC_3$  &$IC_3$  &$ON_2$ &$ER$ &$GR$   &$ACT$\cr
\hline
   &  &\multicolumn{6}{c}{Gaussian population}\cr
\hline
100 &TRUE      &0        &0        &0.1        &4.2      &4.4      &64.3\cr
    &OVER      &0        &0        &0          &6.6      &7.3      &0.10\cr
    &UNDER     &100      &100      &99.9       &89.2     &88.3     &35.6\cr
    &AVE       &1.18     &1        &1.53       &2.29     &2.37     &4.58\cr
\hline
300 &TRUE      &47.0     &1.7      &31.2       &27.0     &28.2     &98.9\cr
    &OVER      &0        &0        &0.1        &0.4      &0.4      &1.1\cr
    &UNDER     &53.0     &98.3     &68.7       &72.6     &71.4     &0\cr
    &AVE       &4.42     &2.81     &4.17       &3.01     &3.07     &5.01\cr
\hline
500 &TRUE      &0        &0        &98.8       &88.9     &89.7     &98.9\cr
    &OVER      &0        &0        &0          &0        &0        &1.1\cr
    &UNDER     &100      &100      &1.2        &11.1     &10.3     &0\cr
    &AVE       &2.44     &1.16     &4.99       &4.76     &4.78     &5.01\cr
\hline
1000&TRUE      &0        &0        &99.9       &99.9     &99.9     &99.1\cr
    &OVER      &0        &0        &0.1        &0        &0        &0.9\cr
    &UNDER     &100      &100      &0          &0.1      &0.1      &0\cr
    &AVE       &1.17     &1        &5          &5        &5        &5.01\cr
\hline
    &  &\multicolumn{6}{c}{Uniform population}\cr
\hline
100 &TRUE     &0         &0       &0.1         &5.0      &5.4      &60.7\cr
    &OVER     &0         &0       &0.1         &8.4      &9.0      &0.4\cr
    &UNDER    &100       &100     &99.8        &86.6     &85.6     &38.9\cr
    &AVE      &1.17      &1       &1.57        &2.38     &2.45     &4.54\cr
\hline
300 &TRUE     &48.4      &1.4     &37.8        &31.7     &33.7     &99.4\cr
    &OVER     &0         &0       &0           &0.3      &0.4      &0.6\cr
    &UNDER    &51.6      &98.6    &62.2        &68.0     &65.9     &0\cr
    &AVE      &4.45      &2.83    &4.27        &3.16     &3.25     &5.01\cr
\hline
500 &TRUE     &0         &0       &99.4        &91.0     &91.6     &99.1\cr
    &OVER     &0         &0       &0.1         &0        &0        &0.9\cr
    &UNDER    &100       &100     &0.5         &9.0      &8.4      &0\cr
    &AVE      &2.44      &1.13    &5           &4.81     &4.83     &5.01\cr
\hline
1000&TRUE     &0         &0       &99.9        &100      &100      &99.0\cr
    &OVER     &0         &0       &0.1         &0        &0        &1.0\cr
    &UNDER    &100       &100     &0           &0        &0        &0\cr
    &AVE      &1.12      &1       &5           &5        &5        &5.01\cr
\hline
\end{tabular}
\end{center}\label{table3}
\end{table}

\begin{table}[htbp]
\caption{Percentages of the estimated number of common factors for Case 3 with $n=300$ in 1000 simulations:
``TRUE", ``OVER" and ``UNDER" truly estimates, overestimates and underestimates the number of common factors, respectively. ``AVE" is the average of the estimated number of common factors.}
\renewcommand{\arraystretch}{0.56}\doublerulesep 0.01pt\tabcolsep 0.18in
\begin{center}
\begin{tabular}{cc|cccccc}
\hline
\hline
$p$&   &$PC_3$ &$IC_3$  &$ON_2$  &$ER$ &$GR$   &$ACT$\cr
\hline
   &  &\multicolumn{6}{c}{Gaussian population}\cr
\hline
100 &TRUE      &0        &0        &0.1       &5.5      &5.8      &0\cr
    &OVER      &0        &0        &0         &9.6      &9.7      &0\cr
    &UNDER     &100      &100      &99.9      &84.9     &84.5     &100\cr
    &AVE       &1        &1        &1.27      &2.51     &2.54     &1.06\cr
\hline
300 &TRUE      &0        &0        &1.1       &4.2      &4.6      &5.4\cr
    &OVER      &0        &0        &0         &0.8      &0.9      &0\cr
    &UNDER     &100      &100      &98.9      &95       &94.5     &94.6\cr
    &AVE       &1        &1        &2.85      &2.1      &2.14     &2.91\cr
\hline
500 &TRUE      &0        &0        &32.5      &26.0     &27.3     &71.3\cr
    &OVER      &0        &0        &0         &0.2      &0.2      &2.8\cr
    &UNDER     &100      &100      &67.5      &73.8     &72.5     &25.9\cr
    &AVE       &1        &1        &4.2       &2.92     &2.97     &4.74\cr
\hline
1000&TRUE      &0        &0        &99.6      &92.3     &92.7     &96.2\cr
    &OVER      &0        &0        &0         &0        &0        &3.8 \cr
    &UNDER     &100      &100      &0.4       &7.7      &7.3      &0   \cr
    &AVE       &1        &1        &5         &4.81     &4.83     &5.04\cr
\hline
    &  &\multicolumn{6}{c}{Uniform population}\cr
\hline
100 &TRUE      &0        &0        &0         &5.0      &5.1      &0\cr
    &OVER      &0        &0        &0         &6.8      &7.0      &0\cr
    &UNDER     &100      &100      &100       &88.2     &87.9     &100\cr
    &AVE       &1        &1        &1.27      &2.33     &2.35     &1.08\cr
\hline
300 &TRUE      &0        &0        &0.5       &5.2      &5.5      &4.6\cr
    &OVER      &0        &0        &0         &1.2      &1.3      &0.10\cr
    &UNDER     &100      &100      &99.5      &93.6     &93.2     &95.30\cr
    &AVE       &1        &1        &2.87      &2.25     &2.28     &2.92\cr
\hline
500 &TRUE      &0        &0        &37.3      &31.5     &32.6     &76.10\cr
    &OVER      &0        &0        &0.1       &0.2      &0.2      &1.10\cr
    &UNDER     &100      &100      &62.6      &68.3     &67.2     &22.80\cr
    &AVE       &1        &1        &4.26      &3.08     &3.13     &4.76\cr
\hline
1000&TRUE      &0        &0        &99.8      &94.5     &94.7     &96.8\cr
    &OVER      &0        &0        &0.1       &0        &0        &3.2\cr
    &UNDER     &100      &100      &0.1       &5.5      &5.3      &0\cr
    &AVE       &1        &1        &5         &4.88     &4.88     &5.03\cr
\hline
\end{tabular}
\end{center}\label{table4}
\end{table}

\begin{table}[htbp]
\caption{Percentages of the estimated number of common factors for Case $4$ with $n=300$ in 1000 simulations:
``TRUE", ``OVER" and ``UNDER" truly estimates, overestimates and underestimates the number of common factors, respectively. ``AVE" is the average of the estimated number of common factors.}
\renewcommand{\arraystretch}{0.56}\doublerulesep 0.01pt\tabcolsep 0.18in
\begin{center}
\begin{tabular}{cc|cccccc}
\hline
\hline
$p$&   &$PC_3$  &$IC_3$  &$ON_2$  &$ER$ &$GR$   &$ACT$\cr
\hline
   &  &\multicolumn{6}{c}{Gaussian population}\cr
\hline
100 &TRUE      &0.2        &0        &0.7        &3.9     &4.6     &98.20\cr
    &OVER      &0          &0        &0          &1.9     &2.4     &0.20\cr
    &UNDER     &99.8       &100      &99.3       &94.2    &93      &1.60\cr
    &AVE       &2.4        &1        &2.85       &2.14    &2.21    &4.99\cr
\hline
300 &TRUE      &99.5       &81.7     &97.8       &81.6    &83      &99.3\cr
    &OVER      &0.1        &0        &0.1        &0       &0       &0.7\cr
    &UNDER     &0.4        &18.3     &2.1        &18.4    &17.0    &0\cr
    &AVE       &5          &4.81     &4.98       &4.55    &4.6     &5.01\cr
\hline
500 &TRUE      &63.9       &18.5     &100        &99.9    &99.9    &99.4\cr
    &OVER      &0          &0        &0          &0       &0       &0.6\cr
    &UNDER     &36.1       &81.5     &0          &0.1     &0.1     &0\cr
    &AVE       &4.63       &3.81     &5          &5       &5       &5.01\cr
\hline
1000&TRUE      &4.9        &0.1      &99.9       &100     &100     &99.5\cr
    &OVER      &0          &0        &0.1        &0       &0       &0.5\cr
    &UNDER     &95.1       &99.9     &0.0        &0       &0       &0\cr
    &AVE       &3.6        &2.54     &5          &5       &5       &5\cr
\hline
    &  &\multicolumn{6}{c}{Uniform population}\cr
\hline
100 &TRUE      &0.3        &0        &1.3        &4.7     &5.0     &96.0\cr
    &OVER      &0          &0        &0          &2.4     &2.8     &0.5\cr
    &UNDER     &99.7       &100      &98.7       &92.9    &92.2    &3.5\cr
    &AVE       &2.32       &1        &2.87       &2.21    &2.28    &4.97\cr
\hline
300 &TRUE      &99.6       &88.1     &98.8       &87.7    &88.7    &99.6\cr
    &OVER      &0          &0        &0.1        &0       &0       &0.4\cr
    &UNDER     &0.4        &11.9     &1.1        &12.3    &11.3    &0\cr
    &AVE       &5          &4.88     &4.99       &4.73    &4.76    &5\cr
\hline
500 &TRUE      &67.1       &18.1     &99.8       &99.8    &99.8    &99.7\cr
    &OVER      &0          &0        &0.2        &0       &0       &0.3\cr
    &UNDER     &32.9       &81.9     &0          &0.2     &0.2     &0\cr
    &AVE       &4.66       &3.85     &5          &5       &5       &5\cr
\hline
1000&TRUE      &6.4        &0.2      &99.9       &100     &100     &99.3\cr
    &OVER      &0          &0        &0.1        &0       &0       &0.7\cr
    &UNDER     &93.6       &99.8     &0          &0       &0       &0\cr
    &AVE       &3.71       &2.54     &5          &5       &5       &5.01\cr
\hline
\end{tabular}
\end{center}\label{table5}
\end{table}

\section{Empirical Studies}\label{ES}
This section analyzes two real data on economics and finance to demonstrate our proposed estimation method ACT.

{\bf Example 1} (Macroeconomic time series): We use the monthly macroeconomic datasets from March, 1960 to December, 2014 used by \cite{MichNg}. Series 64, 66, 101 and 130 are removed because of missing observations.
Following \cite{MichNg}, outliers are removed where an outlier is defined as an observation that deviates from the sample mean by more than ten interquantile ranges.
After the datasets are cleaned, the data dimension is $p=123$ and the sample size is $n=583$. \cite{MichNg} used $PC_2$ to select nine factors by using the sample covariance matrix
whose nine largest eigenvalues are $3.91\times10^{11}$, $1.20\times10^{10}$, $4.77\times10^{9}$, $3.06\times10^{9}$, $7.25\times10^8$,
$3.60\times10^8$, $1.38\times10^{8}$ and $2.83\times10^7$ and $9.00\times10^6$. However, the marginal variances of these 123 time series vary widely from $2.80\times10^{-4}$ to $1.80\times10^{11}$, which jeopardizes the fidelity of the covariance matrix based methods.
Our estimation method ACT selects six factors by using the sample correlation matrix whose top nine eigenvalues are
$73.10$, $17.81$, $10.22$, $7.00$, $4.80$, $1.97$, $1.53$, $1.17$, $1.06$.

In terms percent of variance explained by the selected factors, the 9 selected factors explain 99.99\% of total variation, whereas the 6 selected factors explain 99.95\% of total variation.  This is mainly due to the leading eigenvalue which is an order of magnitude larger than the rest.
If we look at the standardized variables (the eigenvalues from the correlation matrix), the selected 9 factors explain 96.49\% of total variations whereas the selected 6 factors explain 93.43\%.

We now examine whether the number of factors that influences the equity market has changed before and after financial crisis.  As an illustration, we use the stationarily transformed macroeconomic time series \citep{MichNg}

{\bf Transformed Macro Data Before the Financial Crisis:}

We now use the stationarily transformed \citep{MichNg} monthly macroeconomic datasets from January, 1960 to December, 2007 with sample size $n=576$ and $p=123$. Using $PC_2$ as in \cite{MichNg},  nine factors are selected. The nine largest eigenvalues for the sample covariance are $2.28\times10^4$, $13.06$, $1.53$, $0.88$, $0.74$, $0.40$, $0.32$, $0.24$, $0.18$.  Again, the marginal sample variances for these $123$ transformed series vary widely from $8.47\times10^{-7}$ to $2.28\times10^{4}$.  On the other hand,
our estimation method ACT selects 10 factors by using the sample correlation matrix.  The 10 largest eigenvalues of the sample correlation matrix are $18.37$, $8.55$, $7.66$, $6.16$, $5.86$, $4.01$, $3.76$, $3.53$, $2.89$, $2.56$.  The variances explained by 9 selected and 10 selected factors are both around 99.99\% due to a very spike top eigenvalue.  In terms of percentage of variance explained by the standardized variables, ten factors explain 51.53\% whereas nine factors explain 49.45\%.

{\bf Transformed Macro Data After the Financial Crisis:}

The period covers the data from January, 2010 to October, 2018 with the sample size is $n=106$ and $p=123$. Again, $PC_2$ selects 9 factors.  The 9 largest eigenvalues are $5.68\times10^4$, $2.77$, $0.86$, $0.35$, $0.17$, $0.11$, $0.08$, $0.08$, $0.03$. Again, the marginal sample variances vary largely from $5.35\times10^{-7}$ to $5.74\times10^{4}$.
In contrast, our estimation method ACT chooses 7 factors.  The 9 largest eigenvalues of correlation matrix are $16.48$, $12.20$, $9.23$, $5.75$, $5.68$, $5.27$, $4.22$, $3.82$ and $3.53$.  
Moreover, nine selected factors explain 53.83\% total variation in 123 series, whereas 7 factors explain 47.85\% total
variation in 123 series, which is similar to pre-crisis period by using the same method.

{\bf Example 2} (100 Fama-French portfolios):
We now estimate the number of factors using the excess returns of Fama-French 100 portfolios. The data can be downloaded from the data library of Professor Kenneth French's website.  Again, we divide the data into two periods:  before and after financial crisis.

{\bf Before the Financial Crisis:}

Following \cite{FanZhangYu2012}, we use the daily returns of 100 industrial portfolios formed on the basis of size and book-to-market ratio from January 2, 1998 to December 31, 2007.  We note that the $71$th and $100$th portfolios have very large variances that possibly jeopardize the covariance matrix based methods.  $PC_3$, $IC_3$, $ON_2$, $ER$ and $GR$ estimate the number of factors as $10$, $10$, $6$, $3$ and $3$, respectively.  The largest 10 eigenvalues of the sample covariance matrix are $1824.45$, $885.13$, $117.39$, $9.74$, $5.38$, $3.17$, $2.31$, $2.14$, $1.86$, $1.59$.
Ten factors explain $98.86\%$ total variation in the $100$ portfolios; four factors (suggested by ACT) explain $98.29\%$ total variation in $100$ portfolios. On the other hand,
ACT selects four factors.  The largest 10 eigenvalues of sample correlation matrix are $65.81$, $5.74$, $2.57$, $1.95$, $1.10$, $0.97$, $0.90$, $0.83$, $0.72$, $0.63$.
Ten factors explain $81.26\%$ total variation, whereas four factors explain $76.09\%$ total variation in $100$ portfolios, in the standardized variables.

\begin{table}[htbp]
\caption{Percent $R^2$ of well-known risk factors explained by PC-factors}
\begin{center}
\begin{tabular}{c|cccc}
\hline
               &Rm-Rf  &SMB     & HML     & Momentum \cr
Before crisis (4 selected factors)  & 0.953 &  0.931 &  0.829  &  0.141 \cr
Before crisis (3 selected factors)&  0.947  & 0.813  & 0.821 & 0.132 \cr
After crisis (3 selected factors)   & 0.982 &  0.891 &  0.917  &  0.155 \cr
\hline
\end{tabular}
\end{center}\label{table6}
\end{table}

The well-known risk factors for equity markets are Fama-French factors
\citep{fama1993common,fama2015five} and  the  momentum factor \citep{carhart1997persistence}.  To examine how these known factors can be explained by the unsupervised learning method (PCA with number of factors selected by ACT),  we regress each known risk factor on the four principal component factors and report the coefficients of determination $R^2$ in Table~\ref{table6}.  As comparison, we also regress these Fama-French factors on the 3 selected factors and report the result in the same table.

First of all, the well-known Fama-French factors (Rm-Rf is the market factor; SMB is the size factor; HML is the value factor; these three factors are nearly uncorrelated) are explained very well by the factors learned from the principal components.  Regarding PC-factors as true factors (subject to learning or estimation errors), the results lend further support that the Fama-French factors are three most important factors, spanning essentially the same space as the first three PC-factors (regressing on first 3 PCs yields similar results).  Such a confirmation of Fama-French factors appears new.  On the other hand, the momentum factors can not be explained well by the first four principal components, which is a surprise.  Figure 1 depicts how well these four well-known equity risk factors can be explained by the four principal components.  As expected, the four principal components explain the Fama-French factors better than the first four principal components.  On the other hand, the Carhart's momentum factor is not supported by the principal components.

\begin{figure}
\centerline{
\begin{minipage}{0.35\textwidth}
\centerline{\includegraphics[height=4cm, width=4cm]{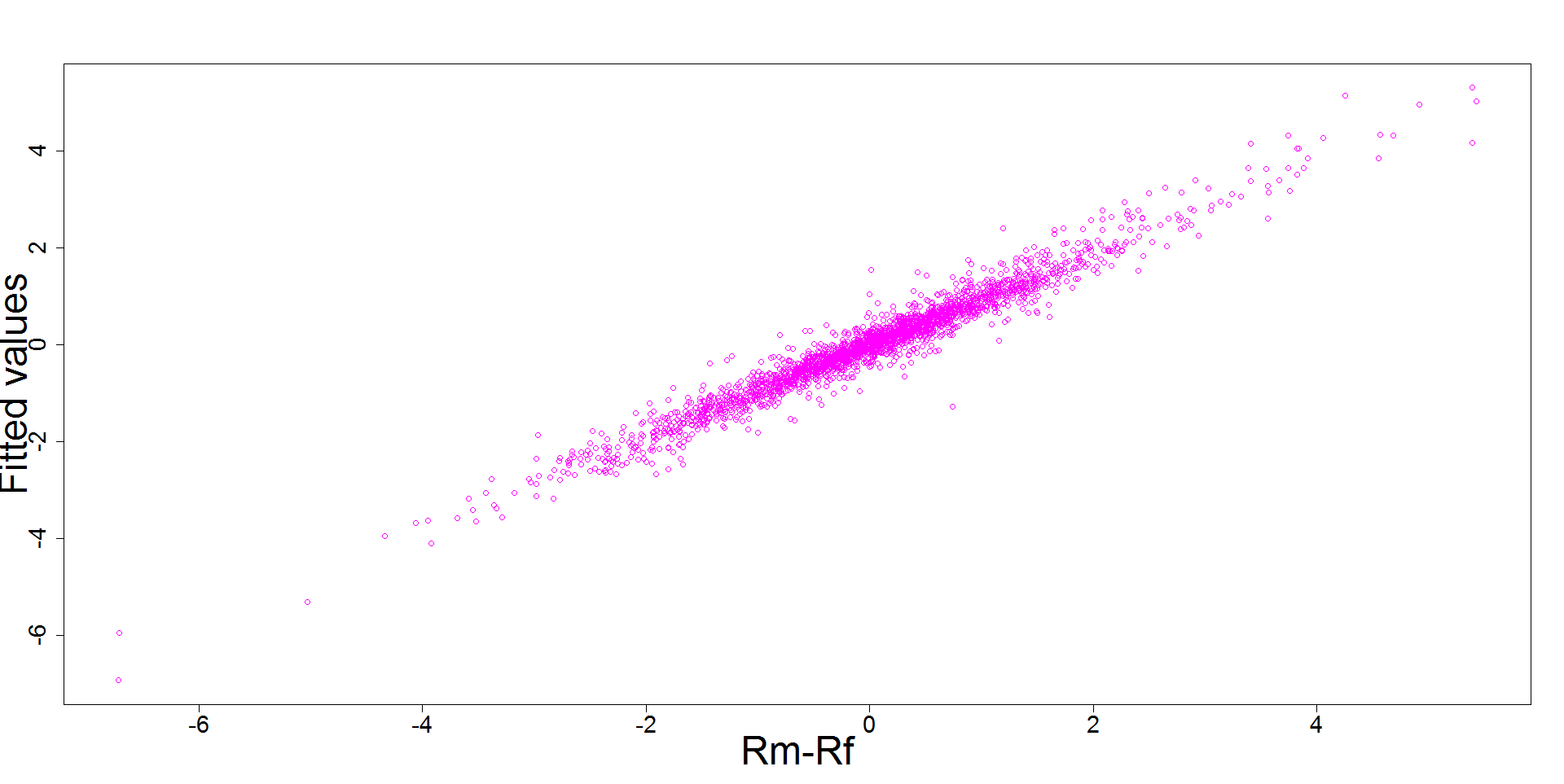}}
\end{minipage}
\begin{minipage}{0.35\textwidth}
\centerline{\includegraphics[height=4cm,width=4cm]{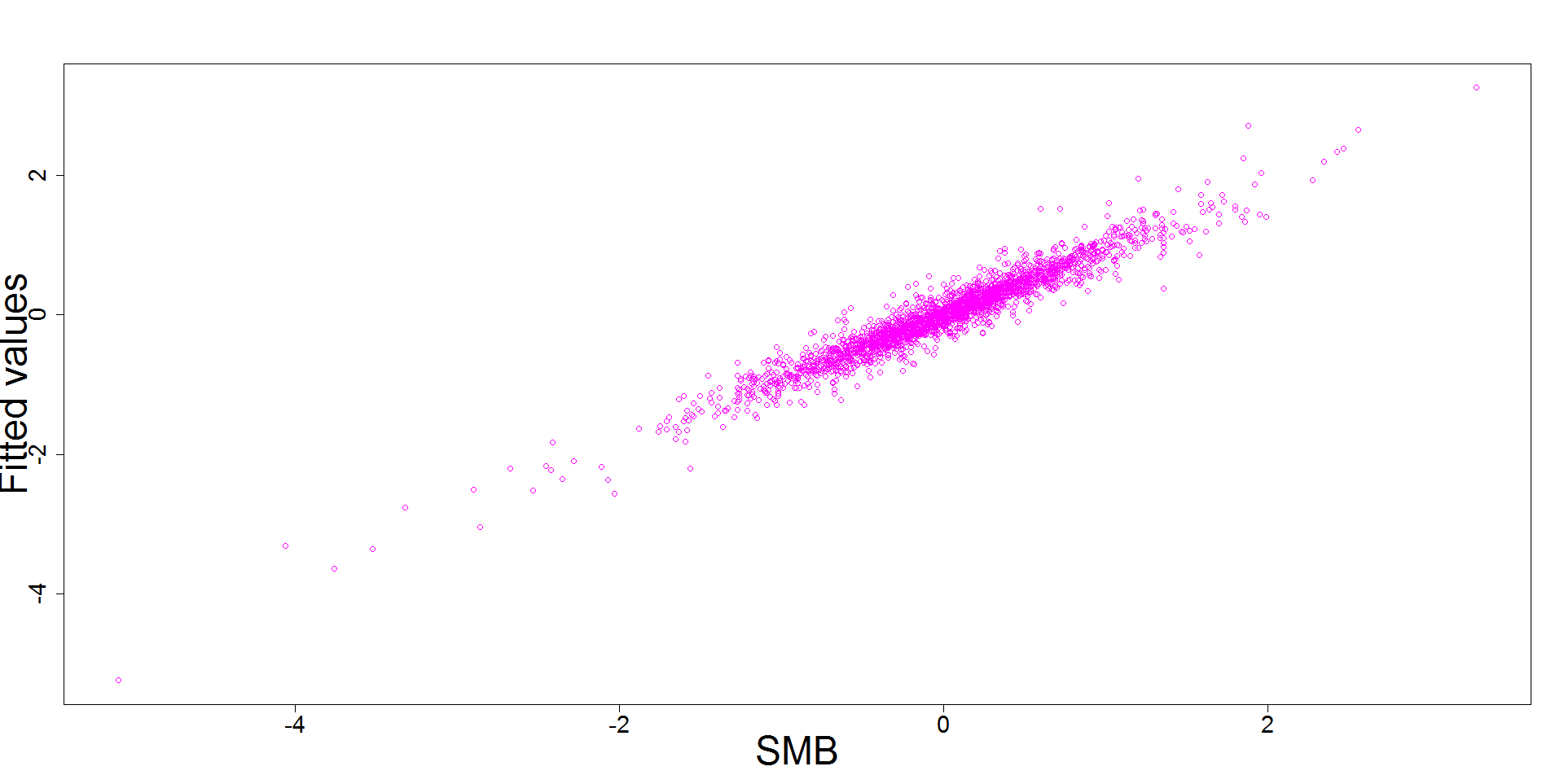}}
\end{minipage}
         }
 \centerline{
\begin{minipage}{0.35\textwidth}
\centerline{\includegraphics[height=4cm, width=4cm]{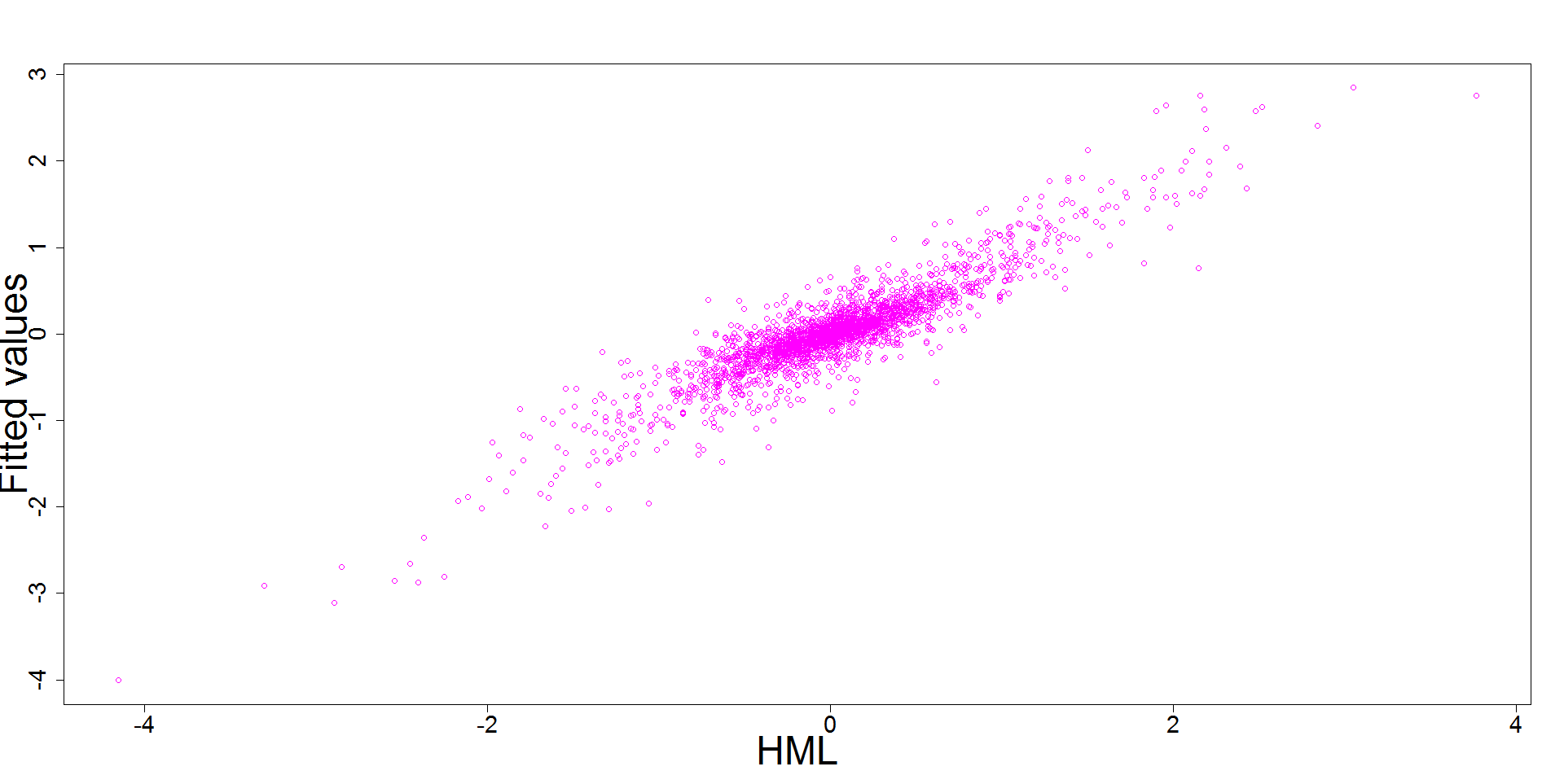}}
\end{minipage}
\begin{minipage}{0.35\textwidth}
\centerline{\includegraphics[height=4cm,width=4cm]{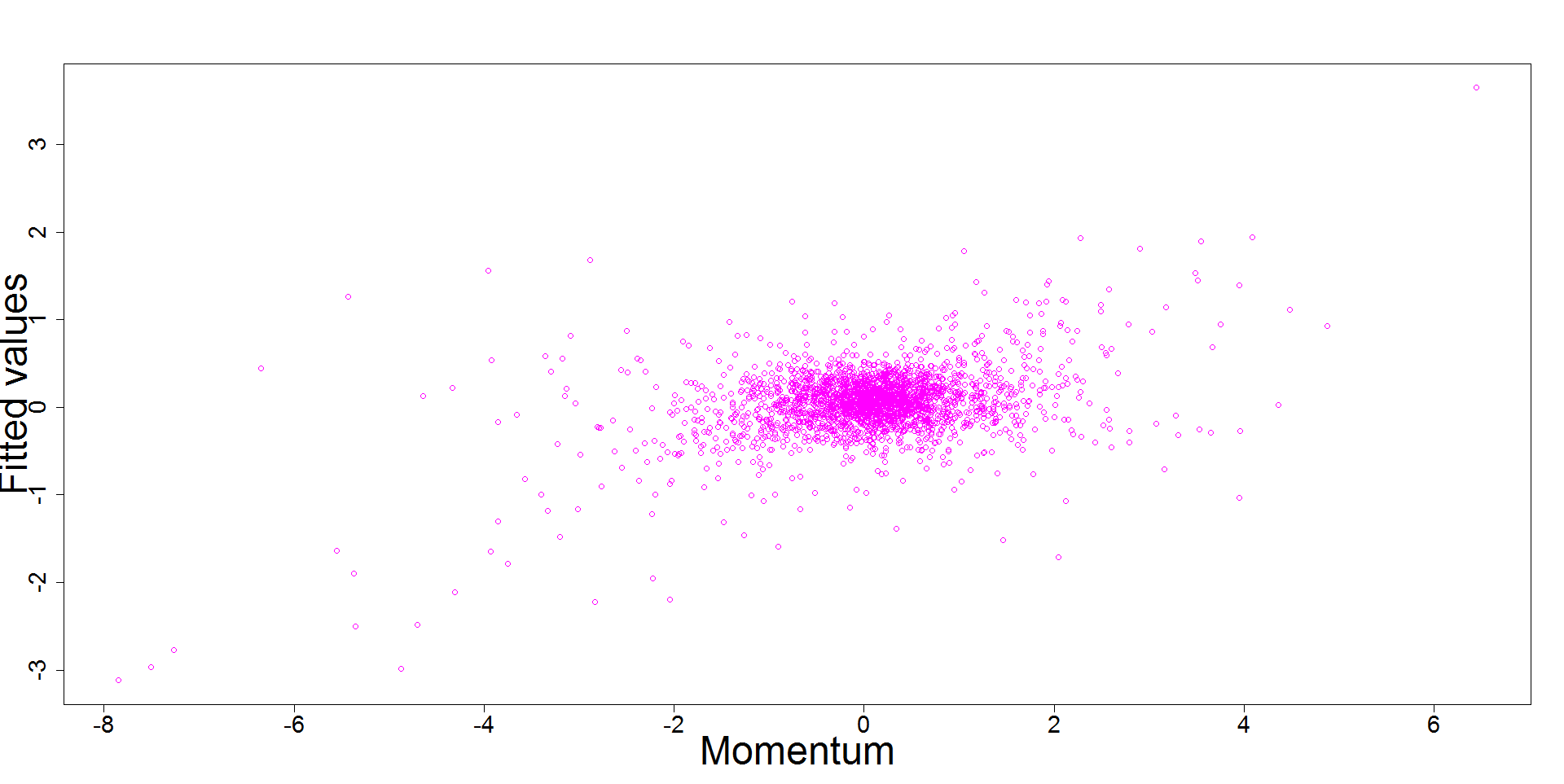}}
\end{minipage}
           }
\caption{Graphical display for every observable factor v.s. its fitted values by regressing every observable factor on the four principal component factors before economic crisis.}
\label{fig1}
\end{figure}

We measure the difference between four given risk factors and four learned factors by using their projection matrix.
Let $A$ be a $n\times 4$ matrix formed by the time series of the four known factors and $B$ be an $n\times 4$ matrix formed the four principal component factors.  Define the projection matrix as $P_A=A (A^TA)^{-1} A^T$ and $P_B=B (B^TB)^{-1} B^T$.  We then measure the difference between the space spanned by the four well-known factors and four learned PC-factors by using the Operator norm and Frobenius norm.  For our data, they are
     $$\|P_A - P_B\|_2=0.973,\quad \|P_A - P_B\|_F=1.591.$$
In a similar vein, we measure the difference between the 3 Fama-French factors (A) and the three principal factors (B) by using the projection matrices.  They are
$$
\|P_A-P_B\|_2 = 0.481, \quad \|P_A-P_B\|_F=0.949.
$$

{\bf After the Financial Crisis:}

We extrac the data from January 4, 2010 to April 30, 2019.
The covariance matrix based methods $PC_3$, $IC_3$, $ON_2$, $ER$ and $GR$ estimate number of factors as $6$, $6$, $6$, $1$ and $1$, respectively.  On the other hand,
ACT selects three factors which explain $85.90\%$ total variation in $100$ portfolios with three largest eigenvalues being $80.62$, $3.22$ and $2.06$ based on the sample correlation matrix.

The $R^2$ of each the four well-known risk factors determined by three principal component factors is depicted in Table~\ref{table6}.  Again, this confirms once more that the famous Fama-French factors aligned well with the first 3 principal components.  Indeed, the differences between the two spaces are
     $$\|P_A - P_B\|_2=0.406,\quad\|P_A - P_B\|_F=0.708,$$
smaller than what it is before the financial crisis.  On the other hand, the momentum factors are still not explained well by the PC factors.

\section{Conclusions}\label{Conclusion}
Based on the sample correlation matrix, this paper discovers the equality between the number of eigenvalues exceeding one and the number of latent factors.  To utilize such a relationship, we study the random matrix theory based on the sample correlation in order to correct the biases in estimating the top eigenvalues and to take into account of estimation errors in eigenvalue estimation.  This gives rise naturally to the  adjusted correlation thresholding (ACT) for determining the number of common factors in high dimensional factor models.  The estimation method overcomes the disadvantages of using the sample covariance matrix which allows observable variables incomparable in their scales.
Simulation studies show that our proposed estimation method outperforms competing methods in the literature.
This paper considers the iid samples from the static factor model. But in practice, people also care about the dynamic factor model. Our future work
will establish the relationship between the population correlation matrix and the number of common factors in the dynamic factor models,
and propose estimating the number of factors in the high dimensional dynamic factor model.

\bigskip
\begin{center}
{\large\bf SUPPLEMENTARY MATERIAL}
\end{center}
\begin{description}
\item[Title:] Supplementary material for ``Estimating Number of Factors by Adjusted Eigenvalues Thresholding".
The material includes 8 lemmas and their proofs, and the proofs of Lemma 1 and Theorem 1, 2, 3. (SuppleFileFactor.pdf)
\item[R codes for ACT:] R codes are used for simulation studies in Section~\ref{Simu} and empirical studies in Section~\ref{ES} (simuexam zipped file).
\item[Data sets:] Data sets are used in empirical studies in Section~\ref{ES}. (data zipped file)
\end{description}


\end{document}